\documentclass[12pt,a4paper,onecolumn,fleqn]{article} 

\usepackage[lmargin=.5925in,rmargin=.5925in,tmargin=1.in,bmargin=1in]{geometry}

\usepackage[sort&compress]{natbib}
\bibpunct{}{}{,}{s}{}{,} 

\usepackage{graphicx}
\usepackage{epstopdf}

\usepackage[english]{babel}

\usepackage{mathtools}

\usepackage[amssymb,alsoload=hep]{SIunits}
\usepackage{subfig}

\usepackage{url}
\usepackage[mathcal]{euscript}

\usepackage{authblk} 

\input{journal_abreviations}

\begin{document}


\title{Hot super-Earths stripped by their host stars}

\author[1,2]{M. S. Lundkvist*} \author[1]{H. Kjeldsen}
\author[1]{S. Albrecht} \author[3,1]{G. R. Davies} \author[4]{S. Basu} \author[5,1]{D. Huber} \author[1]{A. B. Justesen} \author[1,6]{C. Karoff} \author[1]{V. Silva Aguirre} \author[1]{V. Van Eylen} \author[1]{C. Vang} 
\author[1]{T. Arentoft} \author[7,8]{T. Barclay} \author[5,1]{T. R. Bedding} \author[3,1]{T. L. Campante} \author[3,1]{W. J. Chaplin} \author[1]{J. Christensen-Dalsgaard} \author[3,1]{Y. P. Elsworth} \author[9]{R. L. Gilliland} \author[1]{R. Handberg} \author[10,1]{S. Hekker} \author[11]{S. D. Kawaler} \author[1]{M. N. Lund} \author[12,1]{T. S. Metcalfe}
\author[3,1]{A. Miglio} \author[7,13]{J. F. Rowe} \author[5,1]{D. Stello} \author[1]{B. Tingley} \author[14,1]{T. R. White}

\affil[1]{Stellar Astrophysics Centre (SAC), Department of Physics and Astronomy, Aarhus University, Ny Munkegade 120, DK-8000 Aarhus C, Denmark, \protect\url{*lundkvist@phys.au.dk}}
\affil[2]{Zentrum f{\"u}r Astronomie der Universit{\"a}t Heidelberg, Landessternwarte,
K{\"o}nigstuhl 12, 69117 Heidelberg, Germany}
\affil[3]{School of Physics and Astronomy, University of Birmingham, Birmingham B15 2TT, UK}
\affil[4]{Department of Astronomy, Yale University, New Haven, CT 06511, USA}
\affil[5]{Sydney Institute for Astronomy (SIfA), School of Physics, University of Sydney, NSW 2006, Australia}
\affil[6]{Department of Geoscience, Aarhus University, H{\o}egh-Guldbergs Gade 2, DK-8000 Aarhus C, Denmark}
\affil[7]{NASA Ames Research Center, Moffett Field, CA 94035, USA}
\affil[8]{Bay Area Environmental Research Institute, 596 1st Street West, Sonoma, CA 95476, USA}
\affil[9]{Center for Exoplanets and Habitable Worlds, The Pennsylvania State University, 525 Davey Lab, University Park, PA 16802, USA}
\affil[10]{Max Planck Institute for Solar System Research, D-37077 G{\"o}ttingen, Germany}
\affil[11]{Department of Physics and Astronomy, Iowa State University, Ames, IA 50011, USA}
\affil[12]{Space Science Institute, Boulder, CO 80301, USA}
\affil[13]{SETI Institute, Mountain View, CA 94043, USA}
\affil[14]{Institut f{\"u}r Astrophysik, Georg-August-Universit{\"a}t G{\"o}ttingen, Friedrich-Hund-Platz 1, D-37077 G{\"o}ttingen, Germany}

\maketitle


\begin{abstract}
\textbf{Simulations predict that hot super-Earth sized exoplanets can have their envelopes stripped by photo-evaporation, which would present itself as a lack of these exoplanets. However, this absence in the exoplanet population has escaped a firm detection. Here we demonstrate, using asteroseismology on a sample of exoplanets and exoplanet candidates observed during the Kepler mission that, while there is an abundance of super-Earth sized exoplanets with low incident fluxes, none are found with high incident fluxes. We do not find any exoplanets with radii between $\mathbf{2.2}$ and $\mathbf{3.8}$ Earth radii with incident flux above $\mathbf{650}$ times the incident flux on Earth. This gap in the population of exoplanets is explained by evaporation of volatile elements and thus supports the predictions. The confirmation of a hot-super-Earth desert caused by evaporation will add an important constraint on simulations of planetary systems, since they must be able to reproduce the dearth of close-in super-Earths.}
\end{abstract}

\noindent
Models predict that the envelopes of exoplanets orbiting close to their host star are stripped by photo-evaporation, which should be evident as an absence of very hot super-Earth sized exoplanets. The simulations by ref. \citenum{ref:lopez2013} show a deficit in the number of exoplanets with radii between $1.8$ and $4 \ R_\oplus$, and that these exoplanets should become comparatively rare for fluxes exceeding $100 \ F_\oplus$ due to photo-evaporation. In addition, the simulations reveal a corresponding increase in the number of rocky planets with $R < 1.8 \ R_\oplus$ caused by the presence of the stripped cores. The existence of a paucity in the radius distribution of close-in exoplanets caused by evaporation is also supported by other theoretical works\citep{ref:owen2013, ref:owen2015, ref:kurokawa2014}.

Previous studies have detected a deficit of exoplanets in the radius-period (or semi major axis) diagram\citep{ref:szabo2011, ref:beauge2013, ref:kurokawa2014}. This so-called sub-Jovian pampas or sub-Jovian desert extends from $3$ to $10 \ R_\oplus$ for periods shorter than $2.5 \ \mathrm{days}$\citep{ref:szabo2011,ref:beauge2013}. However, the absence in the distribution of exoplanets caused by evaporation has escaped a secure confirmation\citep{ref:owen2013, ref:owen2015, ref:rowe2015, ref:howard2012, ref:kurokawa2014, ref:szabo2011, ref:beauge2013} primarily due to uncertain host star parameters. This can now be changed with asteroseismology. Asteroseismology studies the stellar pulsations, and it allows us to determine the properties of many exoplanet host stars to high accuracy\citep{ref:huber2013, ref:chaplin2014_param, ref:campante2015, ref:silvaaguirre2015}, which in turn dramatically improves the planetary properties.

NASA's Kepler mission has provided high-quality data for thousands of potential exoplanets and their host stars\citep{ref:borucki2011, ref:batalha2013, ref:mullally2015, ref:rowe2015}. Here we exploit these data, using asteroseismology, to make a robust detection of the hot-super-Earth desert, a region in the radius-flux diagram completely void of exoplanets. We find that the hot-super-Earth desert is statistically significant and not caused by selection effects or false positives. The detection of the existence of a hot-super-Earth desert confirms that photo-evaporation does play a role in shaping the exoplanet population that we see today. This imposes an important constraint on simulations of the formation and evolution of exoplanetary systems since this effect needs to be taken into account.

\section*{Results}

\begin{figure}[htbp]
\centering 
\includegraphics[width=8.5cm]{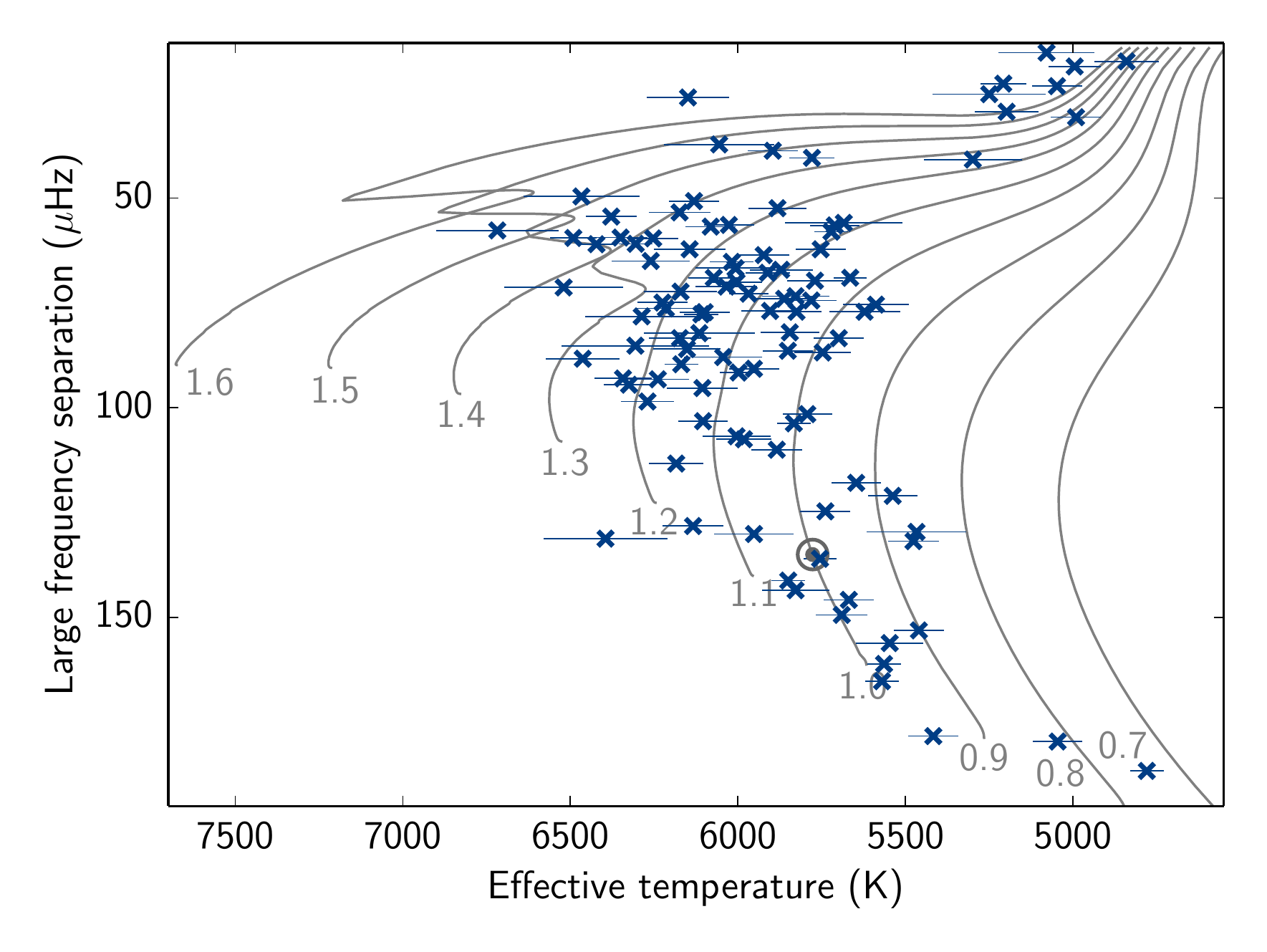}
\caption{\textbf{The seismic host stars.} Asteroseismic Hertzsprung-Russell diagram showing the large frequency separation as a function of stellar effective temperature (with $1\sigma$ uncertainties) for the $102$ exoplanet host stars in the asteroseismic sample. The grey lines show evolution tracks of different masses (in solar units) for solar composition (adapted from ref. \citenum{ref:lundkvist2014}). The location of the Sun is indicated by the grey solar symbol (circle with a dot). \label{fig:1}}
\end{figure}

\noindent
\textbf{The seismic sample of exoplanets}
Using asteroseismology, we obtained accurate stellar mean densities and radii for $102$ exoplanet host stars (both confirmed and candidate exoplanets). These are shown in an asteroseismic Hertzsprung-Russell Diagram in Fig.~\ref{fig:1} (the methods used to determine the parameters are discussed in Methods while Supplementary Table~1 contains the data). The asteroseismic mean densities and radii, combined with precise periods and transit depths as well as the stellar effective temperature, allowed us to calculate very precise planetary radii and incident fluxes for the subset of Kepler exoplanets that orbit the $102$ host stars (typically more precise than $10\%$, see Fig.~\ref{fig:5}).

We determined the flux that the exoplanet receives from its host star, using the following expression for the time-averaged incident flux in units of the Earth value (assuming circular orbits):
\begin{equation}
	\label{eq:incidentflux}
	\frac{F}{F_\oplus} =
		\left( \frac{\rho_*}{\rho_\odot} \right) ^{-2/3}
		\left( \frac{P}{1 \ \mathrm{yr}} \right) ^{-4/3}
		\left( \frac{T_{\mathrm{eff},*}}{T_{\mathrm{eff},\odot}} \right) ^4 \ .
\end{equation}
Here, $\rho_*$ is the stellar mean density obtained from asteroseismology, $P$ is the orbital period, and $T_\mathrm{eff}$ is the effective temperature, with $T_{\mathrm{eff},\odot} = 5778 \ \kelvin$ being the effective temperature of the Sun. To find the radius we used the planet-star radius ratio ($R_\mathrm{p}/R_*$), which can be obtained from the transit depth ($\delta F/F$) and the stellar radius from grid-modelling:
\begin{equation}
	\label{eq:planetradius}
	R_\mathrm{p} = \sqrt{\frac{\delta F}{F}} R_* = \left( \frac{R_\mathrm{p}}{R_*} \right) R_* \ .
\end{equation}
The periods and the planet-star radius ratios have been obtained from ref. \citenum{ref:vaneylen2015} ($62$ exoplanets) or the NASA Exoplanet Archive's cumulative KOI (Kepler Object of Interest) list (\url{http://exoplanetarchive.ipac.caltech.edu/cgi-bin/TblView/nph-tblView?app=ExoTbls&config=cumulative}, accessed on July 1st 2015) with preference given to the former. The uncertainties were estimated using propagation of (Gaussian) uncertainties, where the dominant contribution to the uncertainty on the incident flux stems from the temperature uncertainty.

\begin{figure}[htbp]
\centering
\includegraphics[width=18cm]{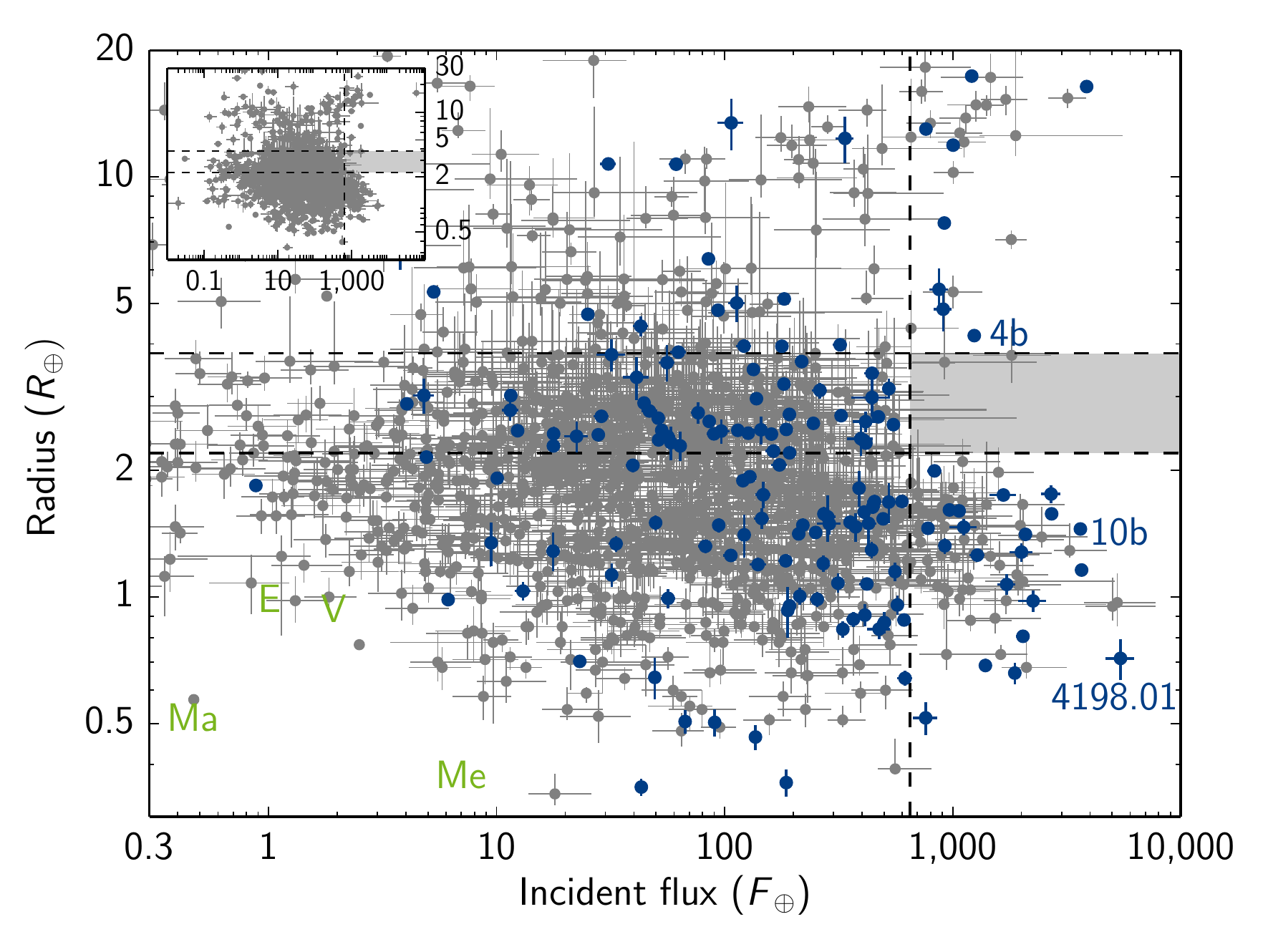}
\caption{\textbf{Radius-flux diagram showing the distribution of exoplanets.} The $157$ exoplanets in the seismic sample are plotted with $1\sigma$ errorbars (in blue), while the non-seismic sample is shown in grey (the inset shows the full non-seismic sample). The exoplanets Kepler-4b, Kepler-10b and KOI-4198.01 are identified in the diagram. The location of the four rocky solar-system planets; Mercury (Me), Venus (V), Earth (E) and Mars (M) is indicated with the green writing (no points). The vertical dashed line marks an incident flux of $650 \ F_\oplus$, while the horizontal dashed lines indicate radii of $2.2$ and $3.8 \ R_\oplus$ respectively. The location of the hot-super-Earth desert has been shaded. \label{fig:2}}
\end{figure}

\vspace{12pt}
\noindent
\textbf{The hot-super-Earth desert}
In Fig.~\ref{fig:2} we show the exoplanet radius as a function of the incident flux for $157$ of the $162$ exoplanets. Five exoplanets were removed from the subsample because their radius estimates had an uncertainty in excess of $20\%$ (in order to not have our sample polluted by bad data points, see Methods for details). For illustration we also show in Fig.~\ref{fig:2} all Kepler KOIs with apparent sizes below $30 \ R_\oplus$ determined to better than $20\%$, that have a calculated flux and are not in our seismic sample (the non-seismic sample, the incident fluxes and radii for these KOIs have been taken from the NASA Exoplanet Archive).

Figure~\ref{fig:2} clearly displays a complete absence of exoplanets with sizes between $2.2$ and $3.8 \ R_\oplus$ and an incident flux above $650$ times the Earth value (shaded area in Fig.~\ref{fig:2}). We constrained the size of the empty region by bootstrapping with $1$~million iterations of the exoplanets present in the seismic subsample and using the boundaries that left the region empty in $95.45\%$ ($2\sigma$) of the iterations (see Methods for further information). This empty region in the radius-flux diagram is the hot-super-Earth desert, and its location agrees with the theoretical prediction\citep{ref:lopez2013}. We note that some data points from the non-seismic sample will fall in the region of the hot-super-Earth desert if no cut is made to weed-out uncertain data points (see for example Fig.~7 of ref. \citenum{ref:rowe2015} and Methods).

As the boundaries and Fig.~\ref{fig:2} suggest, we opted to model the hot-super-Earth desert as a simple box region. While such a simplistic model may not capture the full effects of evaporation on the planet population, we do believe it to encompass the main features. A more sophisticated model taking into account how the amount of evaporation scales with incident flux and planet mass could be a next step.

As an aside, it should be pointed out that in the seismic subsample shown in the radius-flux diagram in Fig.~\ref{fig:2}, KOI-4198.01 (which we call Zenta) appears somewhat isolated. If Zenta is a bona fide exoplanet, it could potentially be a very interesting object, since it has the highest incident flux of the exoplanets in the seismic subsample, and it is below $1 \ R_\oplus$ in size. We have inspected the light curve of Zenta to make sure the transits look like genuine exoplanet transits, and we have obtained a few spectra of the host star with the Nordic Optical Telescope (on La Palma). These spectra will be the subject of a subsequent analysis.

\vspace{12pt}
\noindent
\textbf{Significance}
We employed different techniques to assess the significance of the hot-super-Earth desert. Firstly, under the assumption that the period-, or equivalently the incident flux-distribution does not change with planet radius\citep{ref:morton2014} (the null hypothesis), we tested whether the hot-super-Earth desert could occur by chance. This was done by drawing exoplanets randomly from the planet population below $650 \ F_\oplus$ and counting how many exoplanets fell in the radius range of the desert (see Methods for details). We find that only eight of our $10$~million simulations returned zero exoplanets in the desert. Thus it is very unlikely to observe the desert if the incident flux is not a function of radius, which is in agreement with our observation of a gap in the distribution.

\begin{figure}[htbp]
\centering 
\includegraphics[width=8.5cm]{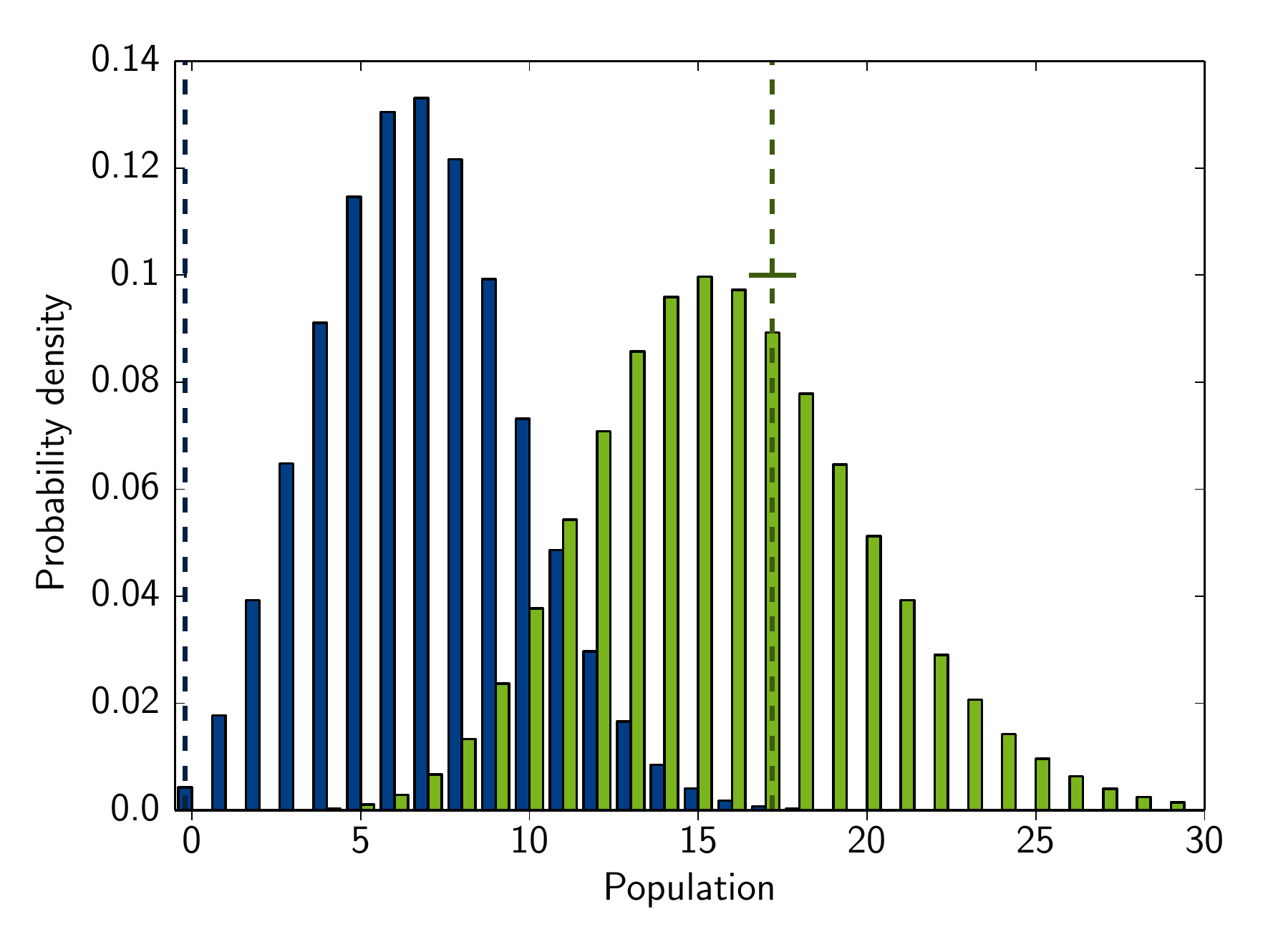}
\caption{\textbf{Simulated number of exoplanets in and below the hot-super-Earth desert.} Histograms of the number of exoplanets in (blue) and below the hot-super-Earth desert (green, in radius) using a Gaussian mixture model of the seismic subsample under the assumption of no gap. The two dashed lines show the observed number of planets in (dark blue) and below the desert (dark green); $0\pm0.04$ and $17\pm0.7$, respectively. The histograms are based on $5$~million realisations. \label{fig:3}}
\end{figure}

Secondly, we used a Gaussian mixture model to represent the seismic subsample as it would look with no desert. Here, the underlying assumption is that the radius-flux distribution can be described by a sum of log-normal distributions\citep{ref:farr2014}. From the model we created a histogram (Fig.~\ref{fig:3}), and we found that fewer than $0.4\%$ of our simulations return the observed number of planets (zero) in the region of the hot-super-Earth desert (see also Methods). This shows that the gap in the radius-flux diagram is significant. It is worth noting that from the non-seismic sample alone, this inference cannot be made (it gives a p-value of $8\%$). We do not believe that this could be due to selection effects between the seismic and the non-seismic sample, since any detected hot-super-Earth planet would have been a high priority target to the Kepler mission. From the Gaussian mixture model treatment of the seismic sample, we also found a slight, although not statistically significant, over-density below the desert (see Fig.~\ref{fig:3}), similar to that expected if the rocky cores are left over from evaporation\citep{ref:lopez2013}.

\vspace{12pt}
\noindent
\textbf{Selection effects and false positives}
It should be emphasized that there are some selection biases in the sample. The limitations in the detection sensitivity of Kepler are the reason for the lack of small exoplanets with low incident flux. Also, for the asteroseismic subsample, the selected stars were on the Kepler short-cadence target list, which is the reason for the low number of large exoplanets with low incident flux (short-cadence slots were prioritised for multi-planet systems over single-planet systems, which favours small planets\citep{ref:latham2011}, and exoplanets showing many transits were discovered early in the mission and kept). While the completeness of the sample is hard to quantify\citep{ref:batalha2013,ref:petigura2013}, no known selection effects\citep{ref:wolfgang2015} would produce the paucity that we observe and the sample is complete down to $2 \ R_\oplus$ for short-period exoplanets\citep{ref:howard2012} (any missing small planets, i.e. below the gap, would only make the desert more pronounced). We also attempted to account for detection biases in our sample by imposing a signal-to-noise ratio criterion and assuming that no exoplanets meeting that criterion with a radius above $1.4 \ R_\oplus$ would have have been missed\citep{ref:huber2013}. We found this not to affect the presence of the hot-super-Earth desert (see Fig.~\ref{fig:7} and Methods for further details).

Despite our basic vetting, the seismic subsample of exoplanets will contain some false positives (FPs). The overall FP-rate for the sample is found to be low\citep{ref:desert2015} (in particular for the multi-planet systems\citep{ref:lissauer2014apj}), but it does vary over the sample. For example, the FP-rate is lower for exoplanets with radii $2-4 \ R_\oplus$ than for those with smaller or larger radii\citep{ref:fressin2013}. However, clearly no FPs have filled the hot-super-Earth desert, and our simulations show we would not be significantly affected by the presence in the sample of the percentage of FPs suggested by ref. \citenum{ref:fressin2013} (see Methods).

\vspace{12pt}
\noindent
\textbf{The sub-Jovian pampas}
A trend, agreeing with our results, has been seen in the radius-period (or semi major axis) diagram by previous studies\citep{ref:szabo2011,ref:beauge2013,ref:kurokawa2014}. They detected a deficit of exoplanets in the radius-range $3-10 \ R_\oplus$ with periods shorter than around $2.5 \ \mathrm{days}$ (the sub-Jovian pampas or sub-Jovian desert)\citep{ref:szabo2011,ref:beauge2013}. While both our hot-super-Earth desert and the sub-Jovian pampas lie at high temperatures (be it high incident flux or short periods), the radius-range is somewhat different, since we find the hot-super-Earth desert to extend only up to $3.8 \ R_\oplus$. Therefore, we investigated the radius-range of the hot-super-Earth desert further to determine whether it could be an extension of the sub-Jovian pampas.

Of the exoplanets in the seismic subsample above our flux boundary of $650 \ F_\oplus$, four exoplanets are present above $10 \ R_\oplus$. These are all confirmed exoplanets (Kepler-1b, 2b, 7b and 14b), and thus agree with the upper limit set by the previous studies. In the radius-range between $4$ and $10 \ R_\oplus$, another four exoplanets are present in our seismic sample. Of these, two are confirmed exoplanets (Kepler-4b and 56b), and a third (KOI-5.01) is a candidate in a multi-planet system (where the FP-rate is lower\citep{ref:lissauer2014apj}). Most important to the location of the upper boundary of the hot-super-Earth desert is Kepler-4b, which is located at $R = 4.2 \ R_\oplus$ and $F = 1243 \ F_\oplus$ (see Fig.~\ref{fig:2}), and thus effectively sets the upper boundary. Kepler-4b has a density of around $1.9 \ \gram \ \centi\meter^{-3}$ (similar to the density of Neptune), and is consequently volatile-rich\citep{ref:borucki2010}, which agrees with its location above the desert. Similarly, Kepler-56b also has a density estimate consistent with a volatile-rich composition\citep{ref:huber2013_sci}, thus also agreeing with its location above the desert.

We have examined the transits of Kepler-4b for evidence that it could be evaporating, but we failed to find any asymmetry in the transits or any transit-to-transit depth variations, which could both indicate atmospheric loss \citep{ref:rappaport2012,ref:rappaport2014}. Still, it cannot be ruled out that evaporation of Kepler-4b could be ongoing at a level, which we cannot detect and possibly at a level that does not influence the radius-evolution of the planet.

In order to assess the significance of an extended gap scenario, we tested two additional sets of boundaries, allowing for the presence of exoplanets from the seismic sample within the gap. The first scenario had the same flux boundary as the hot-super-Earth desert ($650 \ F_\oplus$), but spanned the radius-range from $2.2$ to $10 \ R_\oplus$ in agreement with the upper limit stated for the sub-Jovian pampas. This meant that four seismic exoplanets were present in the tested region (Kepler-4b, Kepler-56b, KOI-5.01 and KOI-1314.01). However, since the sub-Jovian pampas was defined in orbital period rather than incident flux, we cannot replicate the exact limit in the radius-flux diagram found by for instance ref. \citenum{ref:beauge2013}. Thus, we also considered the possibility that we should move the boundary to higher incident flux. Therefore we tested a region with the afore-mentioned limits in radius, but bounded by an incident flux of $1,000 \ F_\oplus$ instead of $650 \ F_\oplus$, which only leaves Kepler-4b in the region of the gap (even though a flux limit this high does not seem to agree with the sharp cut-off in the seismic sample in the $2.2-3.8 \ R_\oplus$ region). We find that both of the tested scenarios are less significant than the hot-super-Earth desert, with the $650 \ F_\oplus$ radius-extended scenario being by far the least significant one.

It can be noted in connection to the sub-Jovian pampas that some exoplanets seem to occupy that gap\citep{ref:beauge2013, ref:colon2015}, and that they do not all appear to be FPs\citep{ref:colon2015}. Ref. \citenum{ref:colon2015} investigate three planet candidates located in the sub-Jovian pampas, and they find that two of the three are likely true planets. While these two planets fall comfortably within the sub-Jovian pampas, one of them is too large to fall in the hot-super-Earth desert, and the other one has uncertainties large enough that it could as well be outside the hot-super-Earth desert (it sits $<1\sigma$ from the upper radius-limit\citep{ref:colon2015}).

\section*{Discussion}

For exoplanets in the radius range in question, radius is thought to be a good proxy for composition\citep{ref:lopez2014,ref:wolfgang2015}. This allows for the transition from a predominantly rocky to a volatile-rich make-up to be expressed in terms of radius, and this transition has been found to occur around $1.6-1.8 \ R_\oplus$ by different studies\citep{ref:rogers2015, ref:lopez2014, ref:wolfgang2015}. Thus, the majority of exoplanets in the $2.2-3.8 \ R_\oplus$ range are expected to be volatile-rich, though some of them could be water worlds\citep{ref:wolfgang2015} (for comparison, the radius of Neptune is $\sim \!\! 3.8 \ R_\oplus$). This agrees with the theory that these exoplanets could be stripped of their envelopes when they are too close to their host star. Thus, we can infer from our hot-super-Earth desert that hot exoplanets below $\sim \!\! 2.2 \ R_\oplus$ most likely have a predominantly rocky composition.

Dynamical interactions may in principle also be responsible for shaping the gap in the radius-flux diagram, for example due to orbital decay or inward migration of planets at late evolutionary stages. However, it seems unlikely that orbital decay played a major part in clearing out the particular part of parameter space associated with the hot-super-Earth desert since the planets would either need to be more massive or on shorter orbits\citep{ref:essick2015}. Other migration channels such as a combination of planet-planet scattering, tidal circularisation and the Kozai mechanism could have played a role in shaping the location of the hot-super-Earth desert through migration of exoplanets that were initially part of a triple (or larger) system\citep{ref:fabrycky2007,ref:nagasawa2008}. These effects have not been considered in our work, but they could be responsible for later migration of some of the planets that sit above the hot-super-Earth desert (and inside the sub-Jovian pampas, such as Kepler-4b). In addition, the flux boundary is likely a function of the planet mass with heavier planets being able to better withstand the evaporation. Therefore, while we find that the hot-super-Earth desert is more significant than the other regions we tested, we are not in a position to unambiguously decide whether the hot-super-Earth desert is an extension of the sub-Jovian pampas or a separate feature in the radius-flux diagram.

We have established the existence of a hot-super-Earth desert in the radius-flux diagram. Its presence confirms that photo-evaporation plays an important role in planetary evolution, with the mass-loss history depending on the incident stellar flux. This represents a mechanism not seen in our own solar system, by which some volatile-rich exoplanets are stripped of their atmospheres by their host stars. Consequently, our detection of a hot-super-Earth desert will add an important constraint for simulations of the evolution of planetary systems.

\section*{Methods}

\noindent\textbf{Preparation of the power spectra}

\begin{figure}[htbp]
\centering 
\includegraphics[width=8.5cm]{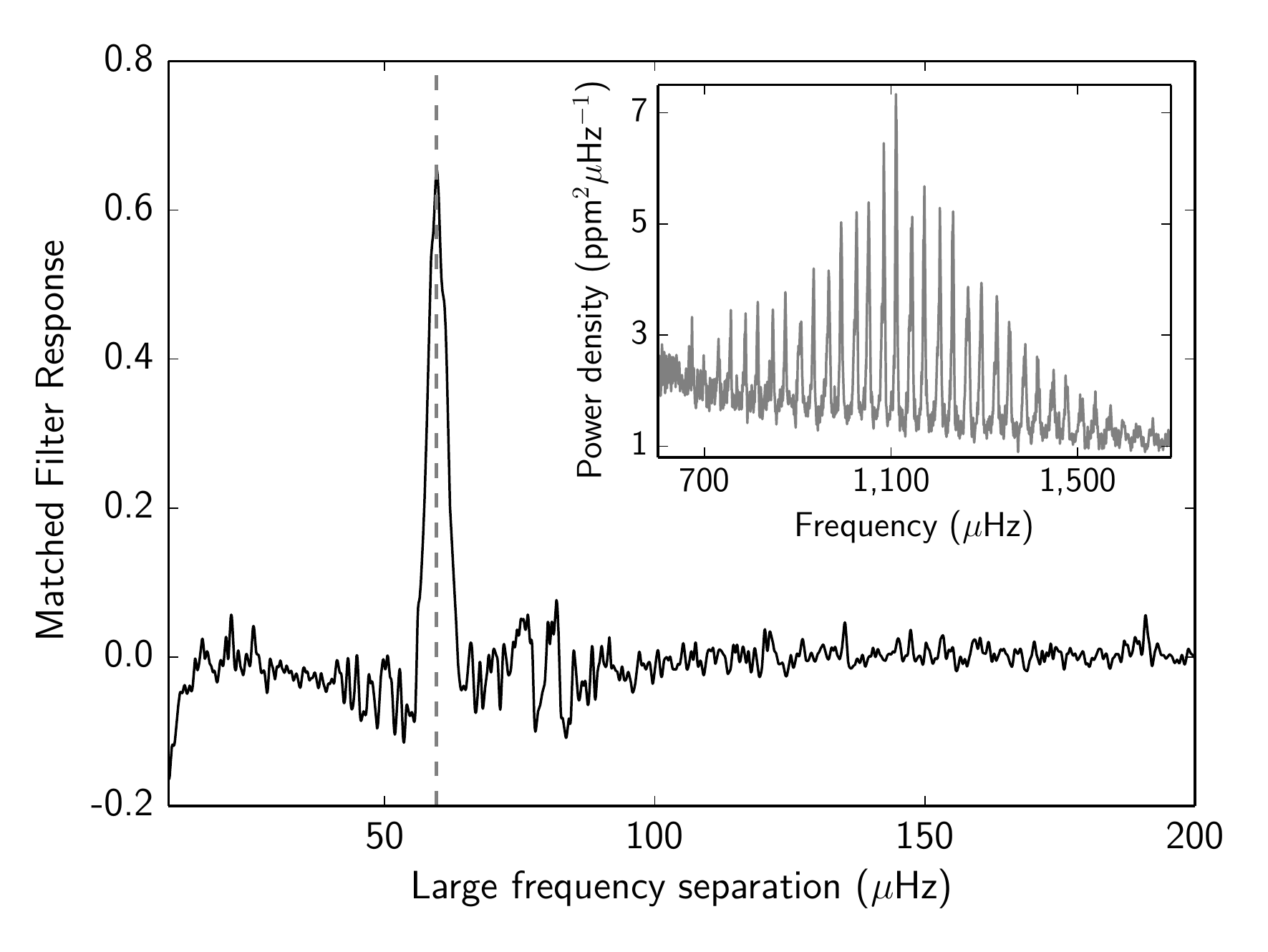}
\caption{\textbf{The large separation.} Output from the Matched Filter Response function for the host star KIC~9414417. The large peak seen at $\sim \!\! 53 \ \micro\hertz$ (and marked by the dashed line) indicates the large frequency separation for the star. The inset shows a section of the power spectrum of KIC~9414417 (smoothed with a Gaussian filter with a width of $1\ \micro\hertz$), where the regular spacing between the oscillation frequencies is clear. \label{fig:4}}
\end{figure}

\noindent Asteroseismology is the study of stellar oscillations. In the case of solar-like stars, the frequencies of the oscillations are almost regularly spaced in a Fourier transform of the time series (a power spectrum, see the inset in Fig.~\ref{fig:4}). The dominant regular structure yields the large frequency separation, which carries information about the stellar mean density\citep{ref:chaplin2013}.

We have searched all $275$ exoplanet host stars with a Kepler magnitude brighter than $13.5$ and with short cadence Kepler data (sampled every $58.85 \ \second$) for an asteroseismic signal. A magnitude limit of $13.5$ was chosen since we have essentially no chance of detecting oscillations in a solar-like star fainter than this\citep{ref:chaplin2011}. To be able to search for the large frequency separation ($\Delta\nu$) for each of the stars, we first made weighted power spectra. The power spectrum for each star was calculated in the following manner: (1) The time series for each quarter (data from Kepler are divided into quarters of approximately $3$ months duration due to the roll of the spacecraft) was cleaned for bad data points using sigma-clipping (with $4\sigma$) of a high-pass filtered time series (high-pass filter was $7 \ \mathrm{min}$) to take out the effect of all slow variations. (2) Using a high-pass filter, the long-term variation of the noise per data point was estimated and taken as the scatter ($\sigma$). (3) Using $1/\sigma^2$ as the statistical weight per data point, we calculated the power spectrum following ref. \citenum{ref:frandsen1995}. (4) For each quarter we calculated a separate power spectrum, and subsequently we combined the power spectra for all quarters into one single spectrum using a weighted mean. The weights were given by $1/\left(\mathrm{median}\left(\mathrm{power}\right)\right)^2$, where the median of the power between $2$ and $4 \ \milli\hertz$ was used. This serves the purpose of down-weighting power spectra for quarters with higher noise levels with respect to the others. Also, when combining several power spectra this way, we change the statistics of the power spectrum from being described by a $\chi_2^2$ to approaching a normal distribution (as stated by the central limit theorem)\citep{ref:appourchaux2003}. An example of a part of a power spectrum can be seen in the inset in Fig.~\ref{fig:4}.

\vspace{0.4cm}
\noindent\textbf{Extraction of large frequency separations}

\noindent A clear asteroseismic signature was found in $102$ of the host stars using a matched filter response function (MFR)\citep{ref:gilliland2011} to search for the large frequency separation. The method takes advantage of the near-regular spacing of the high-order, low-degree p-modes in the power spectrum of solar-like stars. It does this by summing the smoothed power at specific frequencies, which have been calculated from the asymptotic relation \citep{ref:tassoul1980} in the version:
\begin{equation}
	\label{eq:asymp_rel}
	\nu_{n, \ell} \approx \Delta \nu \left( n + \tfrac{\ell}{2} + \varepsilon \right)
								- \ell \left( \ell + 1 \right) D_0 \ .
\end{equation}
Here, $n$ is the radial order of the mode (related to the number of nodes in the radial direction), $\ell$ is the degree of the mode (the number of surface nodes), $\varepsilon$ is a parameter sensitive to the near-surface layers of the star, while $D_0$ is sensitive to the conditions near the core.

When summing the power at frequencies given by different values of $\Delta\nu$ (collapsing over different values of the other parameters in expression~(\ref{eq:asymp_rel}), the result is the MFR giving the summed power as a function of $\Delta\nu$ (see ref. \citenum{ref:gilliland2011} for details). An example for the host star KIC~9414417 can be seen in Fig.~\ref{fig:4}. The large frequency separation corresponding to the most prominent peak in the MFR is then the large frequency separation of the star. The uncertainty on the large frequency separation is determined as the full width at half maximum of the peak.

\vspace{0.4cm}
\noindent\textbf{Grid-modelling of the host stars}

\noindent We used four pipelines to determine the stellar parameters for the $102$ exoplanet host stars. These were Asteroseismology Made Easy (AME)\citep{ref:lundkvist2014}, SEEK\citep{ref:quirion2010}, BAyesian STellar Algorithm (BASTA)\citep{ref:silvaaguirre2015} and the Yale-Birmingham (YB)\citep{ref:basu2010,ref:gai2011,ref:basu2012} pipelines. The YB pipeline derived the properties from five different grids of stellar models, which brings us to a total of eight different grids of stellar models. These pipelines have been used extensively for asteroseismology\citep{ref:huber2013,ref:chaplin2014_param,ref:campante2015}, and further description of the pipelines can be found in the literature.

As inputs to the grid-modelling we used for each star its large frequency separation ($\Delta\nu$) found from asteroseismology and two spectroscopic inputs; the effective temperature ($T_\mathrm{eff}$) and the metallicity ($[\mathrm{Fe}/\mathrm{H}]$). The values that were used for the $102$ host stars can be found in Supplementary Table~1.

We chose to use the mean density and radius returned by AME and then determined the uncertainty by adding in quadrature the uncertainty returned by AME and the scatter over the values returned by the other seven grids. Three stars were too massive for the AME grid, so for these we used the median parameters from the other seven pipelines and estimated the uncertainties by adding in quadrature the median formal uncertainty and the scatter over all seven grids. We note that the parameters returned by the various pipelines were consistent.

Many of the host stars in our seismic sample are present in other large host star samples with published seismic results\citep{ref:huber2013,ref:silvaaguirre2015}, and we have compared the densities and radii obtained for our sample with these other results. We find our parameters to be fully consistent (within $1\sigma$) with the results from ref. \citenum{ref:huber2013} with the exception of Kepler-22. However this is due to the fact that we are using a very different large frequency separation, since the signal originally found\citep{ref:borucki2012} is no longer thought to be the correct one [H. Kjeldsen et al. (in prep.)]. When comparing the densities and radii for the host stars that we have in common with ref. \citenum{ref:silvaaguirre2015} ($32$ stars), we find that all densities and $29$ of the $32$ radii are consistent within $1\sigma$ with the remaining three radii differing by just above $1\sigma$, leading us to conclude that our densities and radii are in agreement with those previously determined.

\vspace{0.4cm}
\noindent\textbf{Vetting of the seismic subsample of exoplanets}

\noindent In order to do some basic vetting of our seismic subsample, we chose to limit our sample to exoplanets that had an uncertainty in radius of less than $20\%$. A large uncertainty on radius was primarily due to large uncertainties on $R_\mathrm{p}/R_*$, which can be caused by grazing transits where the planet only partly covers the star. This removed five exoplanets from the sample: KOI-371.01, KOIs~2612.01 and 2612.02, KOI-3194.01 (which in addition has an impact parameter ($b$\citep{ref:seager2003}) larger than unity) and KOI-5086.01 (also $b>1$). A radius cut of $30\%$ would remove three of these targets (it would leave KOIs~2612.01 and 2612.02 in the sample). It should be noted that none of these exoplanets were situated in the hot-super-Earth desert. Instead of limiting our sample by using the uncertainty in radius, we also tried using the impact parameter (with the criterion $b<1$), which would remove some of the grazing transits. This removed the two exoplanets mentioned above from the asteroseismic subsample, but we opted for the stricter $20\%$ limit on the radius uncertainty.

We also tried to vet the subsample by using asterodensity profiling\citep{ref:vaneylen2015, ref:tingley2011, ref:kipping2014, ref:sliski2014}. Here, the ratio of stellar mean densities derived from the orbit and, in our case, grid-modelling ($\rho_{*,\mathrm{tr}}/\rho_{*,\mathrm{seis}}$) is considered, and a value very different from unity points to either very eccentric orbits or a blend scenario (these are the two largest effects). However, it was difficult to put meaningful constraints on the density ratio since a conservative value did not eliminate any candidates and a more aggressive value would risk throwing away high-eccentricity exoplanets. Thus, we did not pursue this further.

If the cut in radius uncertainty is made at a higher value than $20\%$, exoplanets from the non-seismic sample will appear in the desert. We have examined the points that appear if the cut is instead made at $30\%$ or $40\%$. Using the information from the NASA Exoplanet Archive (from July 1st 2015), when we make the cut at $20\%$ one exoplanet from the non-seismic sample is present in the top of the desert (with its $1\sigma$ errorbars easily placing it outside the desert). If we increase this value to $30\%$, two additional planets enter the hot-super-Earth desert, one very close to the lower flux boundary, and one which, since we downloaded the data, has been flagged as a false positive (FP).

If the cut is instead made at $40\%$, a total of $13$ exoplanet occupy the region of the hot-super-Earth desert including those discussed above. Of these $13$ planets, two are FPs and one is the confirmed exoplanet Kepler-319b. However, upon checking the radius and flux for Kepler-319b listed in the discovery paper\citep{ref:rowe2014}, it is clear that this planet is in fact situated far from the desert (with $R = 1.63 \ R_\oplus$ and $F = 261.6 \ F_\oplus$\citep{ref:rowe2014}), which brings us down to $10$ exoplanets in the desert.

We have manually inspected these $10$ remaining exoplanets situated in the hot-super-Earth desert. They all orbit stars of spectral type F or G, and we find that the reason for the very uncertain exoplanet parameters is very uncertain parameters for the host stars. We find that all of them have uncertainties consistent with a location outside the desert, and that two of them are likely FPs judged on inconsistency between the stellar density derived from the transit and that derived from the stellar mass and radius (these planets are on short orbits and are thus unlikely to have large eccentricities). It is noteworthy that excluding data points with high uncertainties does not exclude a specific spectral type, for instance, it simply limits the number of bad data points in the sample.

\begin{figure}[htbp]
\centering
\includegraphics[width=18cm]{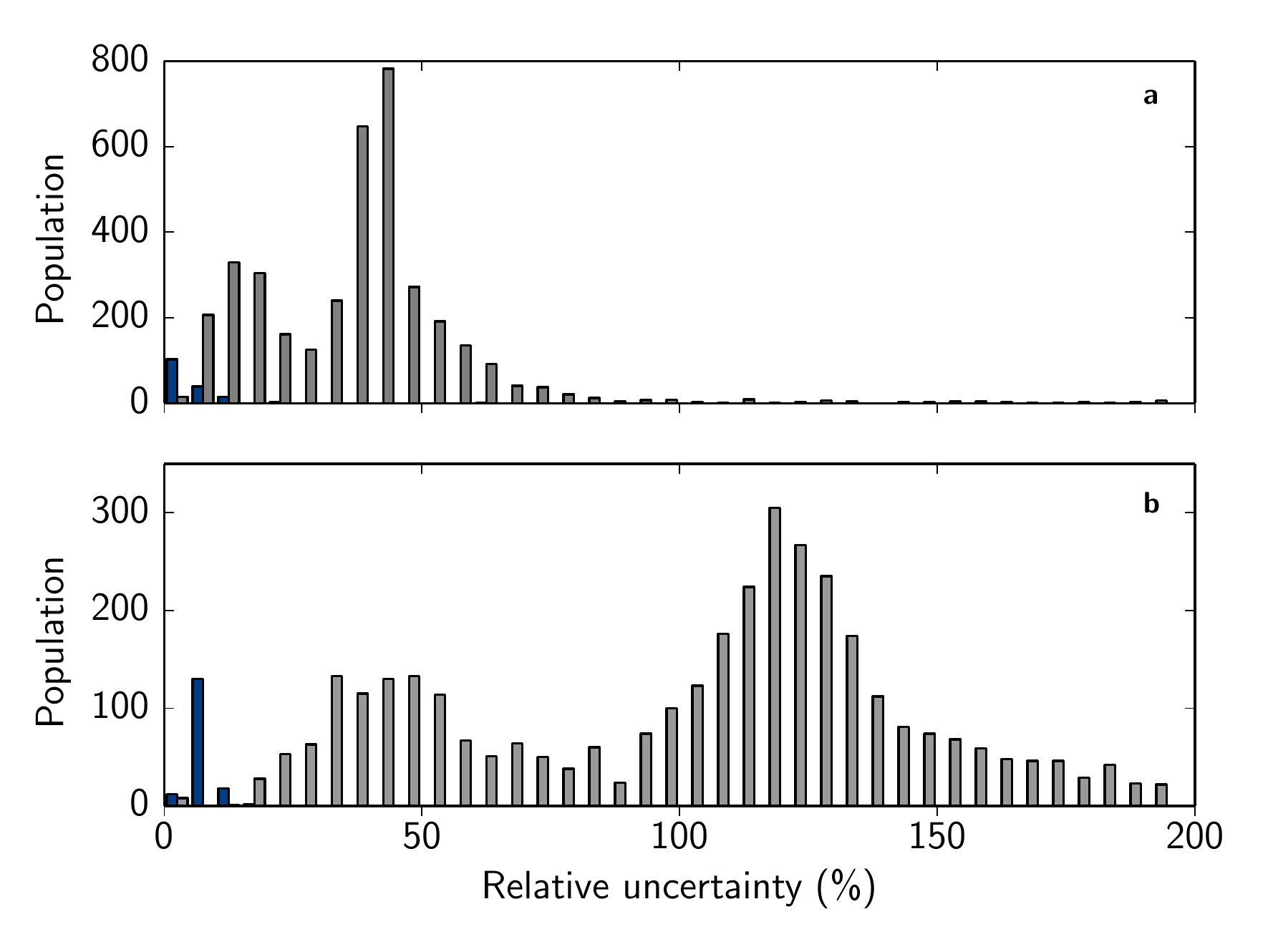}
\caption{\textbf{Distribution of uncertainties on radius and incident flux.} Histogram showing the distribution of the relative uncertainty in radius (\textbf{a}) and incident flux (\textbf{b}) for both the seismic and the non-seismic sample. The seismic sample is shown in blue and the non-seismic sample is shown in grey. Note the bimodal distribution in the non-seismic sample and the low uncertainties in the seismic sample compared to the non-seismic one. \label{fig:5}}
\end{figure}

We have plotted histograms of the relative uncertainties on the radius and incident flux, which can be seen in Fig.~\ref{fig:5}. We note that there is a clear bimodal distribution in both histograms, and that the uncertainties for the seismic subsample are lower than the typical uncertainties in the non-seismic sample. This emphasizes our point that the properties of the seismic sample are determined to a high accuracy. The bimodal distributions in relative uncertainty in flux and radius show that the non-seismic sample is divided into a "low" uncertainty and a "high" uncertainty population, and the division between the two populations lie at $\sim \! \! 30\%$ in radius. Thus, making a cut at $20\%$ should ensure that we are only plotting the best data points from the non-seismic sample, and we have verified that we are not cutting away a population of planets around M-dwarfs (which would have high uncertainties in radius) by doing so.

\vspace{0.4cm}
\noindent\textbf{Determining the boundaries of the hot-super-Earth desert}

\noindent We constrained the size of the hot-super-Earth desert by doing a bootstrap with $1$ million iterations of the exoplanets present in the seismic subsample and using the boundaries that make $95.45\%$ ($2\sigma$) of the iterations return an empty desert. To be specific, we first randomly drew $157$ exoplanets with replacement from the seismic subsample. Then we assigned each of these a radius and a flux randomly selected from Gaussians centred on the parameters for the drawn exoplanet with a standard deviation equal to the uncertainty. Subsequently, we determined how many of these exoplanets that were situated in the hot-super-Earth desert. This was repeated $1$ million times, after which we calculated the percentage of iterations without planets in the hot-super-Earth desert (which is the observed number). We used this information to change the boundaries of the desert, and we repeated the above procedure until we had obtained the $2\sigma$ limits. This procedure does not yield unique boundaries, although they are well constrained due to the small uncertainties on the exoplanets in the sample. However, to determine the exact extent of the hot-super-Earth desert is beyond the scope of this work, and it will in addition depend on whether or not one will allow any exoplanets in the desert.

\vspace{0.4cm}
\noindent\textbf{The Gaussian mixture model}

\noindent  We have used a Gaussian mixture model (GMM), which is a probabilistic model that is the sum of a finite number of Gaussian distributions (we used the Python Scikit-Learn Gaussian Mixture Model\citep{ref:pedregosa2011}). We used the GMM to describe the planet population in log-log radius-flux space and then applied tests to the model to assess the probability that we had detected the hot-super-Earth desert. The distribution of planets in flux and radius is expected to form a correlated log-normal distribution as an outcome of a stochastic planet formation process that produced many correlated, fractional changes in planet sizes and orbits\citep{ref:farr2014}. Thus it is justified to use the GMM, which fit a sum of bivariate Gaussians to the data.

The two different hypotheses that we tested using the GMM and the data are the null hypothesis and the irradiated hypothesis. The null hypothesis states that the radius-flux distribution is smooth, thus that there is no hot-super-Earth desert present in the data. The irradiated hypothesis states that there is a gap in the population density and that there is an over-density at radii lower than the gap.

We leave the number of summed normal distributions as a parameter to be determined by the data in order to allow for different formation processes, selection effects, and other biases. The number of Gaussian components is determined by selecting the model with the lowest Bayesian Information Criterion (BIC). We apply the fit to three different samples; the seismic subsample of exoplanets, the non-seismic subsample and the combined sample. For each sample we use the minimum BIC to determine the number of components used in the GMM. For the seismic subsample, the typical number of components selected by the BIC is one.
 
The fit applied by the GMM does not treat statistical uncertainties on the data points. In order to ensure our tests are robust we have used a Monte Carlo approach to draw each data point from its statistical uncertainties. We generate $1,000$ draws from the uncertainties and for each draw we fit the GMM, and each time we determine the number of components by selecting the lowest BIC. From each of these $1,000$ models, we draw $5,000$ populations and record the number of planets that occupy the gap for each. This then provides the probability distribution of planets in the gap under the null hypothesis, since we fit our model to the data under this assumption.

\begin{figure}[htbp]
\centering
\includegraphics[width=18cm]{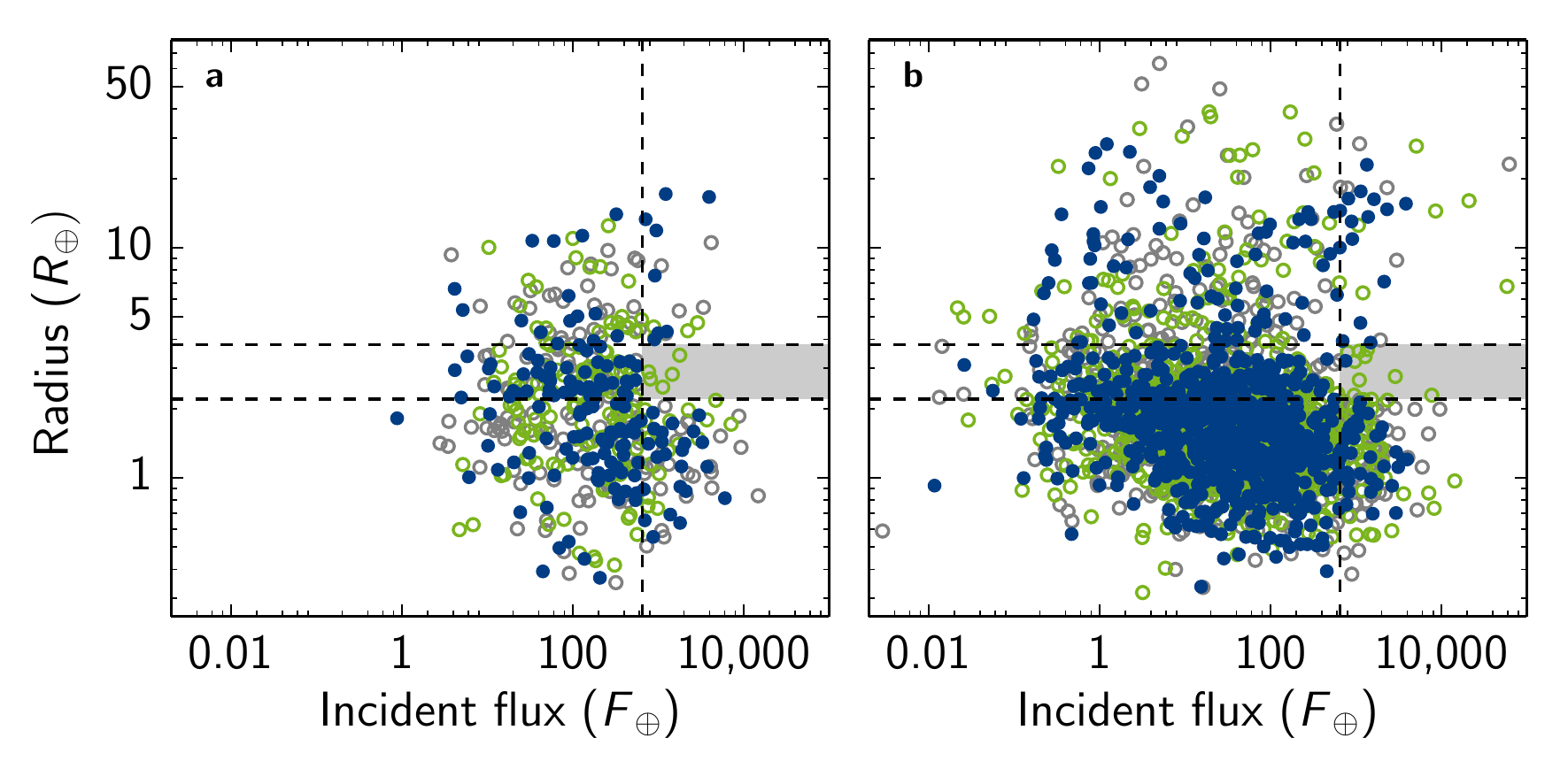}
\caption{\textbf{Radius-flux diagrams showing real and simulated data.} \textbf{a:} One draw of the $157$ exoplanets in the seismic subsample (filled blue cicles) as well as a model of the subsample made from the Gaussian mixture model with (grey open circles) and without an artificial gap (green open circles). The vertical dashed line shows where the incident flux is equal to $650 \ F_\oplus$, while the horizontal dashed lines indicate radii of $2.2$ and $3.8 \ R_\oplus$ respectively. The location of the hot-super-Earth desert has been shaded. \textbf{b:} Same, but for the non-seismic sample. \label{fig:6}}
\end{figure}

Figure~\ref{fig:6} shows examples of the real data together with simulated samples drawn from the fit. We artificially injected a hot-super-Earth desert ($2.2 \leq R_\mathrm{p}/R_\oplus \leq 3.8$ and $F \geq 650 \ F_\oplus$) into the drawn samples by subtracting $2.7 \ R_\oplus$ from the planetary radius if the planet fell within the desert. While somewhat crude, this introduces the gap and an over-density below the gap.

In the seismic sample no planets are observed in the hot-super-Earth desert. Figure~\ref{fig:3} shows the probability distribution of planets expected in the hot-super-Earth desert under the null hypothesis together with the observed value ($0 \pm 0.04$). The uncertainty comes from the small chance that a system actually occupies the gap due to the uncertainties on the planetary radius and flux. Furthermore, we show the expected population distribution below the desert ($0.4 \leq R_\mathrm{p}/R_\oplus \leq 2.2$ and $F \geq 650 \ F_\oplus$) also with the observed value ($17\pm0.7$). The probability of observing no planets in the hot-super-Earth desert in the seismic sample given the null hypothesis is $p = 0.4\%$, which is sufficiently small that we reject the null hypothesis. We observe a slight over-density in the planet population below the desert, but this is very weak and not statistically significant.
 
We repeat the analysis with the non-seismic and the combined samples. For the non-seismic sample we find the probability of observing the data in the desert under the null hypothesis is $p = 8\%$, which supports the rejection of the null hypothesis but is not significant under the typical requirements of either $p < 5\%$ or $p < 1\%$. For the combined data we find a small improvement with $p = 0.3\%$, which is clearly dominated by the seismic sample.

We checked our method using the simulated data with and without a gap. We found results that were consistent with those reported here for the real data.  It should be noted, specifically, that in the simulated-gap seismic sample, we consistently found we could reject the null hypothesis of no gap, while we typically did not confirm an over-density below the gap.  

\vspace{0.4cm}
\noindent\textbf{De-biasing the seismic subsample}

\noindent In an attempt to account for detection biases in our seismic subsample, we de-biased the sample following the approach described by ref. \citenum{ref:huber2013}. First, we determined the minimum planetary radius that should be detectable for a given host star\citep{ref:howard2012}:
\begin{equation}
	\label{eq:rmin}
	R_\mathrm{min} = R_* \left( \sigma_\mathrm{CDPP} \cdot \mathrm{SNR}_\mathrm{lim} \right)^{0.5} \left( \frac{n_\mathrm{tr} t_\mathrm{dur}}{6 \ \mathrm{h}} \right)^{-0.25}  \ .
\end{equation}
Here, $\sigma_\mathrm{CDPP}$ is the $6 \ \mathrm{h}$ Combined Differential Photometry Precision\citep{ref:christiansen2012}, $\mathrm{SNR}_\mathrm{lim}$ the required signal-to-noise ratio (SNR), $n_\mathrm{tr}$ the number of observed transits and $t_\mathrm{dur}$ the duration of a transit. We chose a SNR threshold of $10$ (ref. \citenum{ref:howard2012}), and for each exoplanet in the sample we estimated $R_\mathrm{min}$ by using the median $6 \ \mathrm{h} \ \sigma_\mathrm{CDPP}$ over all observed quarters (obtained from the Mikulski Archive for Space Telescopes, MAST, \url{https://archive.stsci.edu/kepler/}, accessed on July 8th 2015), the transit durations from NASA's Exoplanet Archive's cumulative KOI list (\url{http://exoplanetarchive.ipac.caltech.edu/cgi-bin/TblView/nph-tblView?app=ExoTbls&config=cumulative}, accessed on July 8th 2015), and crudely estimating the number of observed transits by dividing the total lifetime of Kepler (around $1470$ days) by the period of the exoplanet.

\begin{figure}[htbp]
\centering
\includegraphics[width=8.5cm]{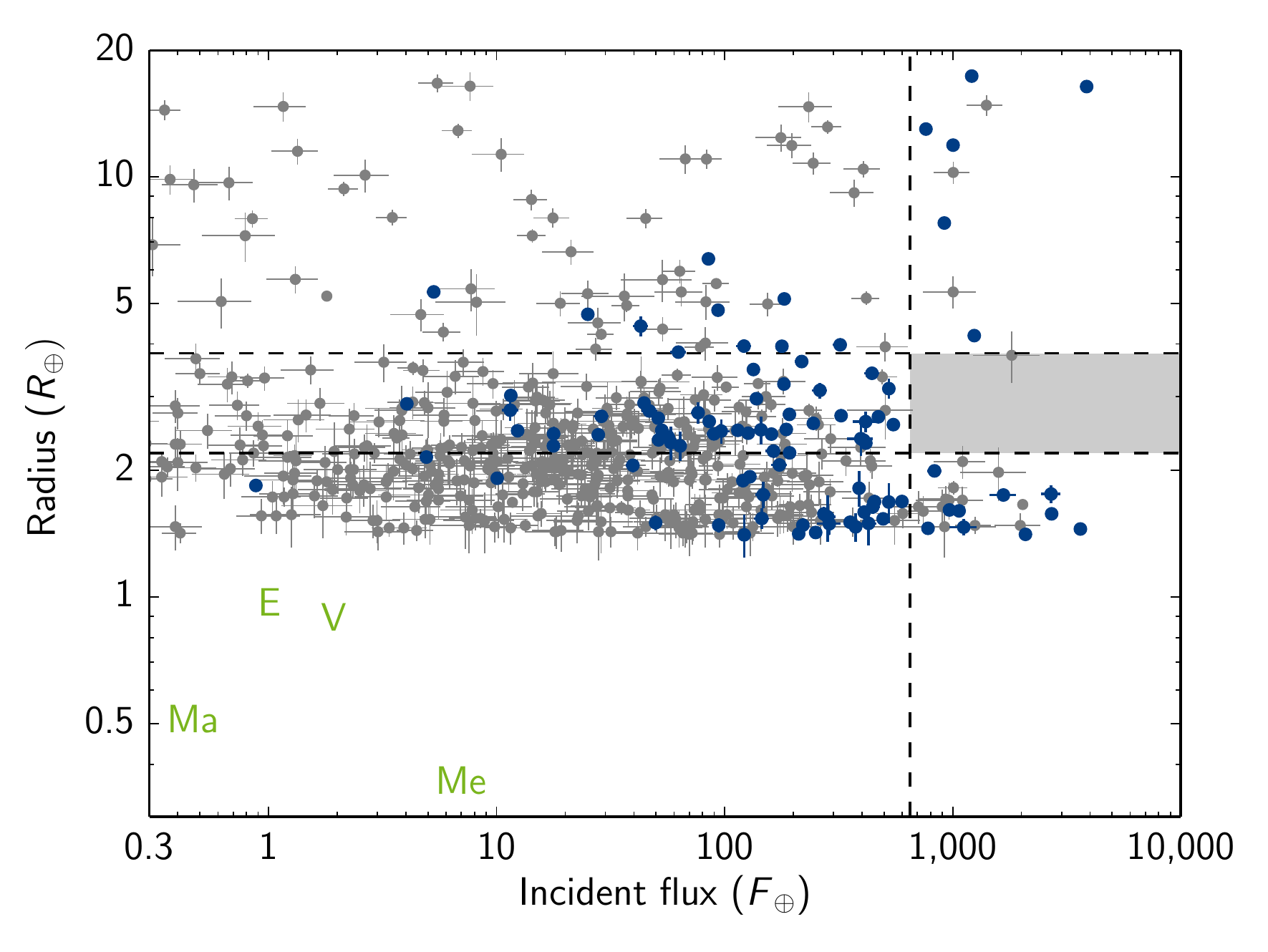}
\caption{\textbf{Radius-flux diagram showing the de-biased seismic subsample.} The de-biased sample of exoplanets with a threshold SNR of $10$ and a radius fulfilling $R_\mathrm{min} < R_\mathrm{x} < R_\mathrm{p}$ with $R_\mathrm{x} = 1.4 \ R_\oplus$ is shown (in blue) with $1\sigma$ errorbars. The grey points are all the exoplanets in the non-seismic sample fulfilling $R_\mathrm{min} < R_\mathrm{x} < R_\mathrm{p}$, also with $1\sigma$ errorbars. The vertical dashed line shows where the incident flux is equal to $650 \ F_\oplus$, while the horizontal dashed lines indicate radii of $2.2$ and $3.8 \ R_\oplus$ respectively. The location of the hot-super-Earth desert has been shaded. \label{fig:7}}
\end{figure}

After having calculated $R_\mathrm{min}$ for all exoplanets in the seismic sample, we found for a range of exoplanet radii ($R_\mathrm{x}$) the number of exoplanets fulfilling the inequality $R_\mathrm{min}<R_\mathrm{x}<R_\mathrm{p}$. Finally, our de-biased sample consists of the exoplanets that fulfil the inequality $R_\mathrm{min} < 1.4 \ R_\oplus < R_\mathrm{p}$, where $1.4 \ R_\oplus$ was the value of $R_\mathrm{x}$ that returned the maximum number of exoplanets. The de-biased sample can be seen in Fig.~\ref{fig:7} along with the de-biased non-seismic sample (for illustration) using the $R_\mathrm{x}$ determined from the seismic subsample. It can be seen that the desert is still evident (also, see below).

\vspace{0.4cm}
\noindent\textbf{Further tests}

\noindent As well as using the Gaussian mixture model to assess the hot-super-Earth desert, we also took another approach to verify the significance of the missing data points. We did this by first dividing the seismic subsample of exoplanets in two groups, with $F/F_\oplus > 650$ and $F/F_\oplus < 650$. We then used the exoplanet radius distribution for the $F/F_\oplus<650$ sample to generate a sample with randomly selected exoplanet radii for a number of exoplanets corresponding to the number of exoplanets with $F/F_\oplus > 650$. Afterwards we determined how many of the selected exoplanets had a radius between $2.2$ and $3.8 \ R_\oplus$. This simulation was repeated $10^7$ times, and we determined the number of times we found the same number of exoplanets as we observe in the hot-super-Earth desert (zero), in this radius range. We found this to happen in only eight of the $10$~million simulations.

This test was repeated to measure the effect of false positives (FPs) on the detection. This was done by randomly removing points from the seismic subsample according to the percentages given by ref. \citenum{ref:fressin2013} and then repeating the above analysis. Of course this approach does not take into consideration any non-uniformity of the FP-rate with flux (for instance there are more eclipsing binaries at higher incident flux\citep{ref:lissauer2014apj}), but on the other hand we do not compensate for the FP-rate for multi-planet systems being lower than for single-planet systems\citep{ref:lissauer2014apj}, or that we have many confirmed exoplanets in the sample (for which the FP-rate should be essentially zero). Thus, this approach should give a fairly conservative estimate of the effect of FPs on the hot-super-Earth desert. We find that $39$ of our $10$~million simulations return zero exoplanets in the radius region of the desert, meaning that the presence of FPs in our sample does not significantly affect the detection of the hot-super-Earth desert.

In order to assess the importance of potential systematic errors on the detection of the hot-super-Earth desert, we investigated the impact on the incident flux of a $100 \ \kelvin$ temperature offset and also of a non-zero eccentricity of $e = 0.5$. We find that the effect of both of these changes is of the same magnitude, and using this test we determined that they have no impact on the detection of the hot-super-Earth desert.

We also used the test on the de-biased sample (seen in Fig.~\ref{fig:7}), and here $54$ of the $10$~million simulations returned the same number of exoplanets in the desert as we observe; thus the detection of the hot-super-Earth desert is not greatly changed by using the de-biased sample instead. We performed a bootstrap on the de-biased sample similarly to what was done for the full sample. Here we found that the boundaries of the hot-super-Earth desert that we determined from bootstrapping the full sample ($2.2 < R_\mathrm{p}/R_\oplus < 3.8$ and $F > 650 \ F_\oplus$) are stronger when considering the de-biased sample. They change from being $2\sigma$ limits to being just above $3.5\sigma$, meaning that $>99.95\%$ of the $1$~million simulations left the hot-super-Earth desert empty ($380$ yielded one planet).

\setlength{\bibsep}{3.0pt plus 0.3ex}

\vspace{0.4cm}
\noindent
\textbf{Author Information}

\vspace{0.4cm}
\noindent
\textbf{Acknowledgements} Funding for the Stellar Astrophysics Centre is provided by The Danish National Research Foundation (Grant agreement no.: DNRF106). The research is supported by the ASTERISK project (ASTERoseismic Investigations with SONG and Kepler) funded by the European Research Council (Grant agreement no.: 267864). This research has made use of NASA's Astrophysics Data System and the NASA Exoplanet Archive, which is operated by the California Institute of Technology, under contract with the National Aeronautics and Space Administration under the Exoplanet Exploration Program.

\vspace{0.4cm}
\noindent
\textbf{Author Contributions} M. S. L. led the work, did grid-modelling, computed the exoplanet parameters, did the bootstrapping, made the de-biased sample and wrote the manuscript. H. K. determined the large frequency separations and performed simulations to assess the significance of the hot-super-Earth desert. M. S. L. and H. K. also designed the project and inspected the exoplanets in the non-seismic sample that fell in the hot-super-Earth desert. S. A. provided feedback on the radius-flux diagram and helped with the structure of the manuscript. G. R. D. ran the Gaussian mixture model and helped with the manuscript. V. V. E. gave feedback on the radius-flux diagram and checked the light curve of KOI~4198.01. D. H. helped with the structure of the manuscript. S. B., C. K. and V. S. A. did grid-modelling. C. V. did an independent analysis of some of the Kepler data. A. B. J. investigated the light curve of Kepler-4b. All authors participated in the interpretation of the results and commented on the manuscript.

\vspace{0.4cm}
\noindent
\textbf{Supplementary information} accompanies this paper at \newline \url{http://www.nature.com/naturecommunications}.
\newline 
\textbf{Competing financial interests:} The authors declare no competing financial interests.


\end{document}